\documentclass[
  aip,
  apl,
  reprint,
  preprintnumbers,
  floatfix,
  amsmath,
  amssymb,
  groupedaddress,
]{revtex4-1}

\usepackage[dvips]{graphicx}
\usepackage{amsmath}
\usepackage{times}

\newcommand{\Fe}{\text{Fe}}
\newcommand{\Cr}{\text{Cr}}
\newcommand{\K}{\text{K}}

\newcommand{\meV}{\text{meV}}
\newcommand{\atom}{\text{atom}}
\newcommand{\Crat}{\text{Cr atom}}
\newcommand{\fig}[1]{Fig.~\ref{#1}}

\begin{document}

\preprint{
  published as Appl. Phys. Lett {\bf 92}, 141904 (2008) \hspace{156pt}
  Copyright (2008) American Institute of Physics
}

\title{
  Are there stable long-range ordered Fe$_{1-x}$Cr$_x$ compounds?
}

\author{Paul Erhart}
\email{erhart1@llnl.gov}
\author{Babak Sadigh}
\author{Alfredo Caro}
\affiliation{
  Lawrence Livermore National Laboratory,
  Chemistry, Materials and Life Sciences Directorate, \\
  Livermore, California, 94550
}

\date{11 March 2008}

\begin{abstract}
The heat of formation of Fe-Cr alloys undergoes an anomalous change of
sign at small Cr concentrations. This observation raises the question
whether there are intermetallic phases present in this composition
range. Here we report the discovery of several long-range ordered
structures that represent ground state phases at zero Kelvin. In
particular we have identified a structure at 3.7\%\ Cr with an
embedding energy which is 49 meV/Cr atom below the solid
solution. This implies there is an effective long-range attractive
interaction between Cr atoms. We propose that the structures found in
this study complete the low temperature-low Cr region of the phase
diagram.
\end{abstract}

\maketitle

Ferritic iron-chromium alloys are important materials for structural
components in fusion and fission reactors.
The need to understand their behavior under irradiation has recently
motivated a significant amount of basic research the results of
which have changed our understanding of these systems.\cite{Hen83,
  MirYalMir04, OlsAbrWal06b, KlaDraFin06}
One of the most surprising and challenging observations is the
coexistence of short-range order and phase segregation in the same 
alloy:
At low Cr-concentrations both experiment \cite{OvcZviLit76,
  MirHenPar84, OvcGolGus06} and simulations \cite{LavDraNgu07,
  ErhCarCar07} have found a short range ordering tendency: Cr atoms
seek to maximize the number of their Fe neighbors. At larger Cr
concentrations, however, phase segregation is observed, i.e. Cr atoms
favor Cr neighbors. Some insight into the origin of this peculiar
behavior has been provided by several first-principles calculations
\cite{Hen83, OlsAbrWal06b, MirYalMir04, KlaDraFin06} which found a
negative heat of formation at low Cr concentrations and an inversion
of the heat of formation above approximately $10\%$ Cr. These
observations raise two important questions:
(\textit{i}) Are there intermetallic phases in the Fe--Cr system which
are missing in the phase diagram as we know it, i.e. is there
long-range order?
(\textit{ii}) Is the Cr--Cr interaction in the Fe host crystal purely
repulsive, i.e. do Cr atoms in Fe behave like a lattice gas, or is
there an effective attraction between the Cr atoms?
In the following we will demonstrate by means of density-functional
theory (DFT) calculations that
(\textit{i}) there are several intermetallic phases at small Cr
concentrations and that
(\textit{ii}) some of these phases correspond to an effective {\em
  long-range attractive} interaction between the Cr atoms.

\begin{figure}[hb!]
  \centering
  \includegraphics[width=0.8\columnwidth]{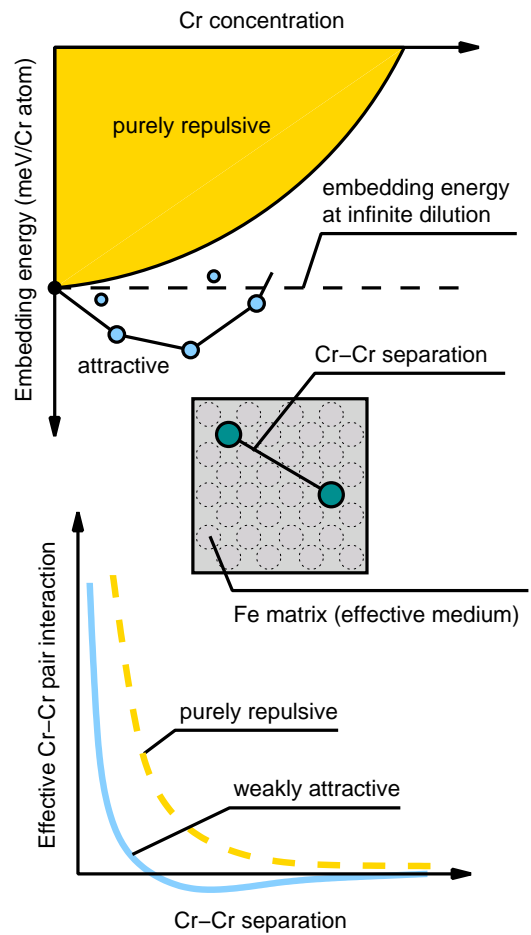}
  \caption{
    (Color online)
    Purely repulsive vs weakly attractive interactions.
    If the Cr--Cr interaction is purely repulsive the embedding energy
    per Cr atom for any possible configuration can only be less
    negative than the embedding energy at infinite dilution.
    If there is, however, some effective attraction between the Cr
    atoms, intermetallic phases exist and there are structures for
    which the embedding energy per Cr atom lies below the infinite
    dilution limit.
  }
  \label{fig:schematic}
\end{figure}

\begin{figure}
  \centering
  \includegraphics[width=0.96\columnwidth]{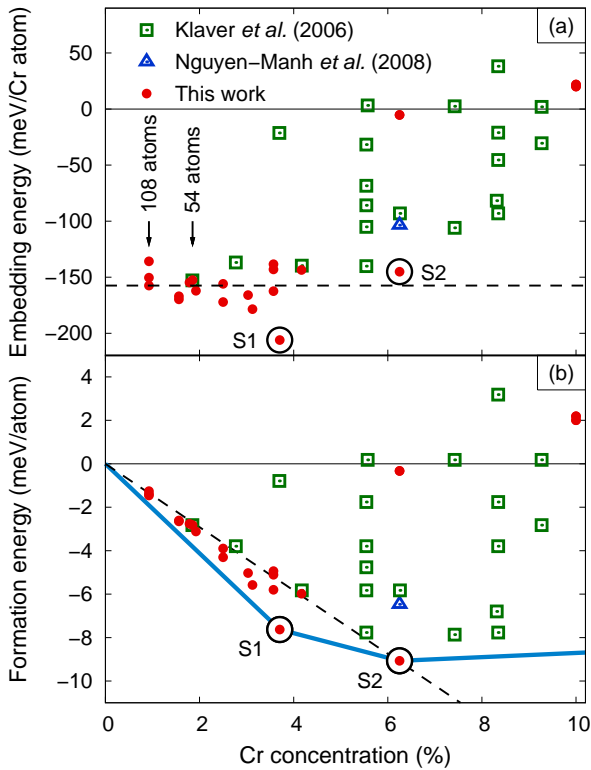}
  \caption{
    (Color online)
    Calculated formation energies plotted (a) as embedding energy
    per Cr atom and (b) formation energy per atom.
    In (a) the structure with the lowest formation energy per Cr atom
    is marked by S1. In this structure the Cr atoms occupy
    a bcc sublattice three times as large as the underlying Fe bcc
    lattice. In (b) the most stable intermetallic phases S1 and
    S2 are connected by thick solid lines. For compositions which
    fall on these lines, the system decomposes into a two-phase
    mixture. For concentrations above 6.25\%\ Cr the thick solid line
    connects S2 with pure Cr.
  }
  \label{fig:results}
\end{figure}

In order to motivate our approach we show in \fig{fig:schematic} a
schematic plot which illustrates the consequences of both purely
repulsive as well as attractive effective interactions between the Cr
atoms. For the following qualitative discussion we treat the Fe matrix
as an effective medium into which the Cr atoms are embedded. The
energy gained if a single Cr atom is removed from a bulk Cr crystal
and inserted into a pure Fe matrix defines the heat of solution for a
single (isolated) impurity, i.e. the {\em embedding energy} at
infinite dilution which is indicated by the horizontal dashed lines in
Figs.~\ref{fig:schematic} and \ref{fig:results}(a). In the case of
purely repulsive Cr-Cr interactions adding any other number of Cr
atoms can only render the embedding energy {\em per Cr atom} less
negative which leads to the shaded region in \fig{fig:schematic}. In
contrast, if attractive forces are present there will be
configurations in which the embedding energy per Cr atom is {\em more}
negative than the dilute limit as indicated by the solid circles in
\fig{fig:schematic}. In either case, since the alloy segregates for
larger amounts of Cr, the embedding energy must eventually become
positive as more and more Cr atoms are added. Finally, it should be
noted that in the case of purely repulsive interactions intermetallic
phases are possible but need not exist. However, if the interaction is
attractive intermetallic phases {\em must} exist.

In alloy thermodynamics, instead of considering the embedding energy,
it is more common to describe a system in terms of the formation energy
(identical to the mixing energy in the case of a random system), which
is the energy {\em per atom} of the alloy compared to the fully
separated components (both Fe and Cr). If one adopts this view the
embedding energy at infinite dilution determines the slope of the
dashed line in \fig{fig:results}(b). If the Cr--Cr interaction is
purely repulsive, all configurations with finite Cr concentration lie
{\em above} this line, whereas at least some phases must lie {\em
  below} this line if there is some attraction.

In several previous studies \cite{KlaDraFin06, LavDraNgu07,
  NguLavDud08a, NguLavDud08b} the formation energies of a number of
Fe--Cr configurations were calculated using supercells derived from
the conventional two-atom body-centered-cubic (bcc) cell (also
compare \fig{fig:results}). On the basis of these data one would
have to conclude that the interaction between Cr atoms is purely
repulsive.


In the present study we have followed an alternative approach. We
systematically generated a large number of structures in which the Cr
atoms selectively occupy certain neighbor shells. No restrictions were
imposed with regard to the symmetry of the Cr sublattice. Specifically
we did not construct our supercells on the basis of the conventional
two-atom body-centered-cubic (bcc) unit cell but considered multiples
of the primitive cell. Any supercell employed in the present work
contains exactly one Cr atom and is uniquely determined by an integer
matrix, $\boldsymbol{A}$, which relates the matrix of primitive cell
vectors, $\boldsymbol{h}$, to the vectors spanning the supercell,
$\boldsymbol{H}$,
\begin{align}
  \boldsymbol{H} &= \boldsymbol{A} \boldsymbol{h}
  \quad
  \text{with}
  \quad
  \boldsymbol{h}=
  \frac{1}{2}
  \left(\begin{matrix}
   -1 &  1 &  1 \\
    1 & -1 &  1 \\
    1 &  1 & -1
  \end{matrix}\right)
\end{align}
where the sum convention applies. Note that the determinant of
$\boldsymbol{A}$ equals the number of atoms in the supercell. Using
appropriate matrices $\boldsymbol{A}$ one can construct supercells
with a variety of different symmetries. Specifically supercells with
face-centered cubic (fcc) and simple cubic (sc) symmetries are
obtained if one uses integer multiples of the matrices
\begin{align}
  \boldsymbol{A_{\text{fcc}}}
  =
  \left(\begin{matrix}
    2 & 1 & 1 \\
    1 & 2 & 1 \\
    1 & 1 & 2
  \end{matrix}\right)
  \quad
  \text{and}
  \quad
  \boldsymbol{A_{\text{sc}}}
  =
  \left(\begin{matrix}
    0 & 1 & 1 \\
    1 & 0 & 1 \\
    1 & 1 & 0
  \end{matrix}\right).
\end{align}

We generated a large set of different supercells focusing on
structures with a depletion of Cr atoms in the first five neighbor
shells. We then employed DFT to calculate the total energies for these
structures \cite{calcdetails} and obtained the formation energies from
\begin{align}
  \Delta E &= E_{\Fe\Cr} - n_{\Fe}/(n_{\Fe}+1) E_{\Fe} - E_{\Cr}
\end{align}
where $n_{\Fe}$ denotes the number of Fe atoms in the supercell,
$E_{\Fe\Cr}$ is the total energy for the supercell,
$E_{\Fe}$ is the total energy of the structurally identical pure Fe
cell, and $E_{\Cr}$ is the energy per atom obtained from a primitive
anti-ferromagnetic Cr cell. If $\Delta E$ is divided by the number of
Cr atoms in the cell one obtains the embedding energy per Cr atom,
$\Delta E_{emb}$, whereas division by the total number of atoms yields
the formation energy, $\Delta E_{f}$.
Special care was taken to minimize the error in our calculations by
varying the cell shape and volume to reduce strain effects. Formation
energies were calculated exclusively as differences between fully
equivalent supercells using fully equivalent $k$-point grids.

The key results of our calculations are compiled in \fig{fig:results}.
The embedding energy at infinite dilution has been obtained using a
108-atom fcc supercell (0.9\%\ Cr). For this structure the neighbor
shells of any Cr atom are depleted of other Cr atoms except for the
25th shell for which 12 out of 36 neighbors are Cr. This corresponds
to a Cr--Cr separation of $4.24\,a_0$ along the $\left<411\right>$
direction where $a_0$ is the bcc lattice parameter. The embedding
energy obtained for this cell is $-157\,\meV/\Crat$ and henceforth is
considered as the infinite dilution limit. For comparison, the lowest
concentration of 1.9\%\ considered in previous studies corresponds to
a structure in which the closest Cr--Cr separation is $3 a_0$ along
the $\left<300\right>$ direction and which yields an embedding energy
of $-153\,\meV/\Crat$.

Our most important discovery is that there are several structures the
energies of which lie below the line given by the infinite
dilution limit. According to the discussion above the location of
these data points implies that there is a net attractive interaction
between Cr atoms.
The structure which shows by far the lowest energy per Cr atom is
marked in \fig{fig:results} by the leftmost black circle (S1). In this
27-atom supercell with 3.7\%\ Cr the Cr atoms are arranged on a bcc
superlattice three times as large as the underlying bcc Fe
lattice. The first nine neighbor shells are exclusively occupied by Fe
atoms whereas the 10th shell is filled to one quarter by Cr atoms. The
Cr--Cr pairs are aligned along the $\left<111\right>$ direction and
separated by $2.60\,a_0$. The embedding energy for this structure is
$-206\,\meV/\Crat$ which is $49\,\meV/\Crat$ below the infinite
dilution limit. There are several other structures at concentrations
smaller than 3.7\%\ Cr with energies below the infinite dilution
limit. They resemble the lowest energy structure in that they display
a depletion of Cr atoms at short range and partly or fully Cr filled
shells at longer range.

For concentrations larger than 3.7\%\ Cr the structure with the lowest
energy (S2 in \fig{fig:results}) is based on a 16-atom base-centered
monoclinic unit cell with lattice vectors
\begin{align}
  \boldsymbol{H}
  =
  \frac{1}{2}
  \left(\begin{matrix}
    -1 &  3 & 3 \\
     3 & -1 & 3 \\
     3 &  3 & 1
  \end{matrix}\right)
  \quad
  \text{and}
  \quad
  \boldsymbol{A}
  =
  \left(\begin{matrix}
    3 & 1 & 1 \\
    1 & 3 & 1 \\
    2 & 2 & 3
  \end{matrix}\right).
\end{align}
The S2 structure has an embedding energy of $-145\,\meV/\Crat$
equivalent to a formation energy of $-9.1\,\meV/\atom$.
In order to describe this structure on the basis of the conventional
two-atom unit cell, one would have to employ a $16\times 16\times 8$
(4096-atom) supercell. The S2 structure described here is different
from the intermetallic structure described in
Refs.~\onlinecite{NguLavDud08a} and \onlinecite{NguLavDud08b} although
both structures correspond to a Cr concentration of 6.25\%. The
structure described in Refs.~\onlinecite{NguLavDud08a} and
\onlinecite{NguLavDud08b}, however, can be  represented in a $2\times 
2\times 2$ (16-atom) cubic supercell in which the Cr atoms are
arranged on a simple cubic lattice with a lattice constant which is
twice as large as the lattice constant of the underlying bcc Fe
matrix. Furthermore, its formation energy is only $-6.5\,\meV/\atom$
(compare the open triangle in \fig{fig:results} and
Ref.~\onlinecite{NguLavDud08b}) and thus not as negative as the value
calculated of $-9\,\meV/\atom$ for the S2 structure.

In order to determine the zero temperature phase diagram we consider
the energy per the total number of atoms (formation energy) as shown
in \fig{fig:results}(b). The infinite dilution limit is indicated by the
dashed line and the stable phases are connected by thick solid
lines, along which the system decomposes into two-phase mixtures.
The two relevant intermetallic ground state phases occur at 3.7\%\
(S1) and 6.25\%\ (S2) corresponding to the two structures described
above.

It is important to address the magnitude of the observed
effect. Even for the most optimal structures the energy gain per atom
amounts to merely $10\,\meV/\atom$ which corresponds to
approximately $110\,\K$. At higher temperatures the configurational entropy
favors the solid solution. Due to the reduced atomic mobility at these
low temperatures direct experimental observation of the ordering may
be challenging. Nevertheless our discovery suggests that the low
temperature phase diagram of Fe--Cr alloys is more complex than
previously thought.

In summary, in this letter we have demonstrated that there are
intermetallic ground state phases at the Fe-rich end of the Fe--Cr
phase diagram with embedding energies as low as $-206\,\meV/\Crat$ and
formation energies as low as $-9\,\meV/\atom$. The observation of
embedding energies below the infinite dilution limit implies that
there are effective long-range attractive interactions between the Cr
atoms which are the result of many-body effects. In fact, attempts to
capture the observed behavior in terms of simple pair interaction
models were unsuccessful. Extensive further study is required to
elucidate the physical origin of this phenomenon.

This work performed under the auspices of the U.S. Department of
Energy by Lawrence Livermore National Laboratory under Contract
DE-AC52-07NA27344 with support from the Laboratory Directed Research
and Development Program. Generous grants of computer time through the
National Energy Research Scientific Computing Center at Lawrence
Berkeley National Laboratory are gratefully acknowledged.


\begin{thebibliography}{16}
\bibitem{calcdetails}
\bibinfo{note}{Calculations were carried out within density-functional
  theory using the Vienna ab-initio simulation package (\textsc{vasp})
  \cite{KreHaf93, KreHaf94, KreFur96a, KreFur96b} with the projector
  augmented plane-wave method.\cite{Blo94, KreJou99} The
  spin-polarized generalized gradient approximation in the
  parameterization by Perdew, Burke, and Ernzerhof was employed to
  represent the exchange-correlation potential. \cite{PerBurErn96} The
  $k$-point grids for Brillouin zone integrations were chosen to
  ensure a total energy convergence of at least $1\,\meV/\Crat$. For
  Fe the $4s$ and $3d$ electrons were included in the valence, and the
  Cr pseudo-potential included the $4s$, $3d$, and $3p$ electrons.
}.
\end{thebibliography}
\end{document}